\documentclass[aip,rsi,amsmath,amssymb,floatfix,superscriptaddress,%
showkeys,%
reprint,%
author-numerical,%
]{revtex4-2}

\usepackage{graphicx}
\usepackage{dcolumn}
\usepackage{bm}
\usepackage[T1]{fontenc}
\usepackage{mathptmx}
\usepackage{textcomp}
\usepackage{multirow}
\usepackage{textcomp} 
\usepackage{xcolor}
\usepackage{array}
\newcolumntype{P}[1]{>{\centering\arraybackslash}p{#1}}

\begin{document}

\title{Combining AFM imaging and elementally resolved spectro-electrochemistry for understanding stability and quality of passive films formed on Alloy 600}

\author{Dominik Dworschak}
\affiliation{Vienna University of Technology, Institute for Applied Physics, Wiedner Hauptstrasse 8-10, A-1040 Vienna, Austria}
\author{Marina Bishara}
\affiliation{Vienna University of Technology, Institute for Applied Physics, Wiedner Hauptstrasse 8-10, A-1040 Vienna, Austria}
\author{Hsiu-Wei Cheng}
\email{hsiu-wei@tuwien.ac.at*}
\affiliation{Vienna University of Technology, Institute for Applied Physics, Wiedner Hauptstrasse 8-10, A-1040 Vienna, Austria}
\author{Markus Valtiner}
\email{markus.valtiner@tuwien.ac.at*}
\affiliation{Vienna University of Technology, Institute for Applied Physics, Wiedner Hauptstrasse 8-10, A-1040 Vienna, Austria}

\begin{abstract}
Understanding elemental corrosion currents and visualizing corroding topographies provide a detailed insight into corrosion mechanisms at the nano-scale. Here, we develop a strategy to understand the elemental composition, corrosion resistivity and local stability of passive materials. Specifically, we utilize a pulse voltammetry approach in a novel electrochemical AFM cell and complement this data by real-time dissolution currents based on spectro-electrochemical online analysis in an ICP-MS flow cell. We study the oxide properties and their protective behaviour, when formed under different applied potentials using alloy 600 as model sample. 
Both AFM and ICP-MS data show that passive films formed on alloy 600 at around +0.3 to +0.4~V in neutral 1 mM NaCl solution are most stable during anodic corrosion at +1.0~V, while AFM further demonstrates that local dissolution occurs, indicating locally varying defect levels in the passive film. 
In combination of both techniques, our approach provide real-time elementally resolved and localized information of passive film quality under corrosive conditions, and it may prove useful for other corroding materials. 
\end{abstract}


\maketitle

\section{Introduction}
Nickel-based alloys (NBAs) show outstanding corrosion resistance under many extreme environments enabling them to be used in broad applications such as steam reactors, aerospace and chemical processing.\cite{Lu.2007b,rosenberg_nickel_1968,rowcliffe_perspectives_2009,henderson_nickel_2004,persaud_characterization_2018}.
It is well-established that the unique corrosion resistance of NBAs such as alloy 600 can only be achieved when the chromium content is more than 10~ \%\cite{Hayes.2006,Lloyd.2004,ebrahimi_role_2015,persaud_review_2016}, which is necessary for the formation of a stable passive/chemically inert chromium oxide enriched thin film (typically 1-2 nm\cite{Maurice.2018}) on the alloy surface.
With its outstanding corrosion inhibition property, such thin films have attracted great attention from researchers \cite{Lu.2007b,Galtayries.2006} to study their passivation mechanism, as well as to characterize its physical structure, chemical composition and electronic properties.\cite{Marcelin.2016,Mito.2009,Tsuchiya.2002} 
While alloy 600 is prone to pitting corrosion \cite{Merola.2019}, under mild conditions passive film breakdown is considered to be the critical step \cite{Frankel.2017} leading to corrosive failure. 
Hence, understanding the properties and anti-corrosion mechanism of the passive film are central for predicting corrosion tendency of NBA in mild environments. 

Up to now, the formation of passive films have been described using models such as the High Field Mechanism (HFM)\cite{Cabrera.1949}, or the Point Defect Model (PDM)\cite{Chao.1981}, which can be used to explain growth kinetics as well as their chemical composition.
These models suggest that changing the conditions under which the passive layer is formed can influence the formation kinetics that leads to grow distinct passive films, and therefore, result in varying corrosion resistance.

The passivity of the chromium oxide thin film is often understood as formation of a dense physical barrier that is limiting the permeability of ions across the film \cite{Marcus.2012} or a change of the electronic structure (band gap) that creates a higher energy barrier for charge exchange at the interface\cite{tranchida2018electronic}.
It is easy to picture when growing passive films under different potentiostatic polarization, the created potential drop across the oxide film can result in a different degree of ion migration/permeability, which directly influences the passive film structure.

However, due to the nature of its extremely small film thickness, characterizing its physicochemical properties during a corrosive process remains challenging. 
Most of the surface sensitive analytical methods are \textit{ex situ} techniques and very often requiring to operate in specific environment such as under ultra high vacuum. 
In this aspect, obtained results from these techniques do not directly reflect on the \textit{in situ} passive film composition and behaviour. 

Consequently, techniques capable of performing real-time study such as scanning probe techniques e.g. Atomic Force Microscopy (AFM) and Scanning Tunneling Microscopy (STM) are more suitable
in studying actively corroding surfaces, which has increasingly gained more impact in recent years. \cite{Chen.2020, Shi2018, Deng2018,Maurice.2007,Maurice.2018} 

Most of the research conducted observed the growth behaviour of passive films e.g. during linear sweep voltammetry (LSV).
In this perspective, the created oxide films are not well-defined and the composition may vary with the varying potential. 
I.e. properties of a passive oxide from a LSV may be influenced by the scan rate and applied solution conditions. 

\begin{figure}[h]
    \centering
    \includegraphics[scale=0.6
    ]{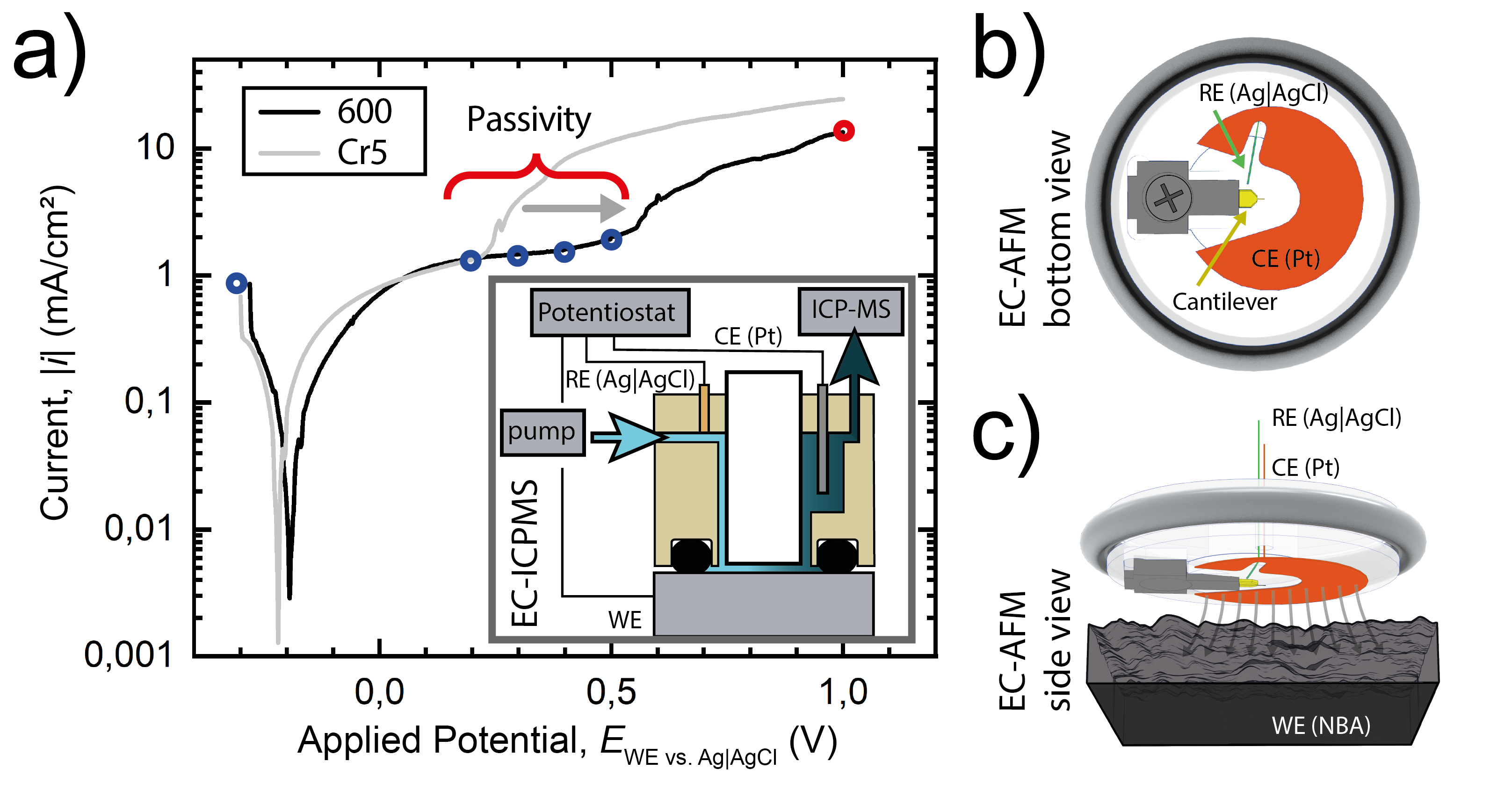}
    \caption
    {
        \textbf{a)} Tafel characteristic curve of the examined alloy: NBA 600 ($Ni_{72}Fe_{8}Cr_{15.5}$) and Cr5 ($Ni_{86}Fe_{9}Cr_{5}$) for comparison. The blue circles mark the regions of interest in this study. 
        Inset shows a schematic of the home-designed ICP-MS coupled electrochemical flow cell. 
        A stamp is used for establishing laminar flow across the surface of working electrode (WE, studied alloy). 
        The counter electrode (CE, Pt) and reference electrode (RE, Ag|AgCl) are arranged to avoid WE being contaminated by the reaction products from CE (hydronium or hydroxide that may vary the pH near WE surface). 
        \textbf{b)} Design principle of the home-designed three-electrode system equipped AFM liquid cell, bottom view. 
        A platinum foil forms the CE directly at the optical window of the used cantilever holder. 
        Only a small section is left out for the AFM laser. 
        Reference electrode (Ag|AgCl) is located outside the cell with a tubing (green) connected. 
        \textbf{c)} Side view of the assembly with the sample (WE) at the bottom in parallel with the CE, resulting in a even electric field distribution.
    } 
\label{fig1}
\end{figure}

It is therefore crucial to develop a strategy to understand the elemental composition, corrosion resistivity and stability of oxide films that formed under \textit{in situ} conditions. 
We considered passive films grown under potentiostatic conditions as most controlled conditions for systematically studying their behaviour.
In this work, we utilize a pulse voltammetry approach in a novel EC-AFM cell (\textbf{Fig. \ref{fig1}b, c}) as well as in a flow cell connected to an inductively coupled plasma mass spectroscopy (ICP-MS) online analysis (inset in \textbf{Fig. \ref{fig1}a}) to study the oxide properties and their protective behaviour, when formed under different applied potentials.
Both techniques provide time-resolved information about the surface condition with the ICP-MS having the focus on elemental composition and the AFM providing topographic/structural information of a materials during corrosive breakdown of alloy 600 as model system. 
Both AFM and ICP-MS data show that passive films formed at around +0.3 - +0.4~V in neutral NaCl solution are most stable during during anodic corrosion at +1.0~V, while AFM further demonstrates that local dissolution occurs, indicating locally varying defect levels in the passive film.

\section{Results and Discussion}
Corrosion resistance of NBA increases with increasing chromium content. 
Exemplary, a Tafel analysis shown in \textbf{Fig. \ref{fig1}a)} for an NBA with 5 \% and 15.5 \% chromium demonstrates the effect of chromium on the window of passivity. 
When comparing the polarization curves, alloy 600 exhibits a well defined region of passivity from -0.1 to +0.5~V shown as the flat plateau in the Tafel plot (indicated by grey arrow), whereas the Cr5 is showing an earlier break-down (light grey) around +0.2~V.

\begin{figure*}
    \centering
    \includegraphics[scale=0.4]{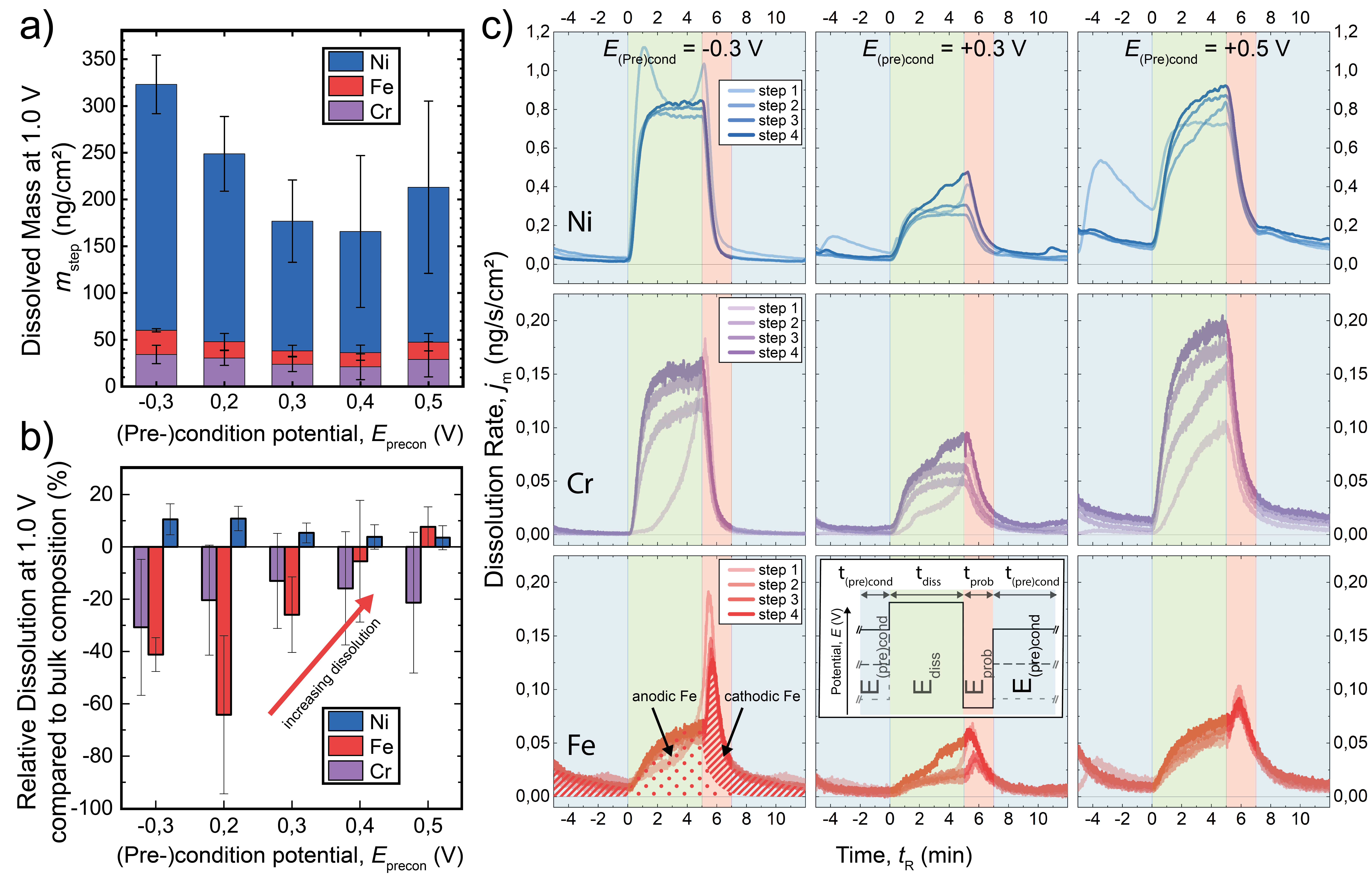}
    \caption{Relative dissolution during anodic polarization.
    Electrochemical flow cell coupled ICP-MS dissolution characteristics in designed passivation-dissolution sequences. 
    \textbf{a)}  
    \textbf{b)} Integration of collected mass during corrosion at 1.0~V over the course of single sequences (4 pulse periods), resolved per elemental contribution.
    Error whiskers show standard deviation of value over three repetitions of the experiment with >4 pulse sequences. 
    \textbf{c)} Overlap of 4 consecutive pulse periods of recorded ICP-MS mass-time profiles for all elements during one full sequence. Please note that the scale for nickel is about 1 order of magnitude higher. 
    Inset shows the schematic of the designed applied electrochemical pulse profile in multi-step amperometry.}
    \label{fig2}
\end{figure*}

\paragraph{Results of ICP-MS experiments.}
The linear polarization shown in \textbf{Fig. \ref{fig1}a)} changes the passive film properties continuously, our interest is however to study potential dependent passive film properties and their protective characteristics. 
Hence, in order to approach the passive film formation under thermodynamic equilibrium, we introduced a multi-step chronoamperometry sequence (see inset figure of \textbf{Fig. \ref{fig2}c)}) to perform step-wise surface oxide growth, corrosion resistance test and reconditioning as shown in \textbf{Fig. \ref{fig2}}.
Specifically, freshly polished alloy 600 was first polarised at a passivation potential of $E_{(pre)cond}$ over 5~minutes for growing a passive layer. 
Subsequently, the polarization jumped to $E_{diss}$ = +1.0~V for another 5~minutes for testing the corrosion resistance of the material after passive film formation at the given $E_{(pre)cond}$.
At the end of the corrosion test, polarization switched back to reconditioning at $E_{recond}$ = -0.3~V for 2~minutes to remove the remaining surface oxides, which resets the alloys' surface chemical state for the coming repetition sequences.

This approach can therefore provide a clear insight into protective characteristics of passive films prepared under different electrochemical potentials.
Based on the Tafel analysis shown in \textbf{Fig. \ref{fig1}a)} we selected five potentials of interest: $E_{(pre)cond}$ = -0.3, +0.2, +0.3, +0.4 and +0.5~V, where -0.3~V was chosen as reference region.
There we expect no or only minor oxide film formation as it is located at the cathodic branch of the Tafel plot and as is estimated from Pourbaix diagrams.\cite{beverskog_revised_1996}
The other potentials are well in the passive region.

We used ICP-MS and EC-AFM (see setup in \textbf{Fig. \ref{fig1}b)} and \textbf{c)}) to identify the alloy's potential-dependent elemental dissolution tendency and surface morphology change, respectively.
The advantage of such flow-cell coupled ICP-MS arrangement is that it provides online corrosion product analysis at trace levels, which can directly be correlated to the applied electrochemical polarization during a real-time process.
In combination with EC-AFM, we may further understand local and micro structure effects.

\textbf{Fig. \ref{fig2}a)} shows integrated dissolved mass of nickel, chromium and iron marked in blue, purple and red, respectively, from ICP-MS during polarization of +1.0~V averaging over four repetitions (see \textbf{Fig. \ref{fig2}c)}).
Among all the examined potentials, we surprisingly found that alloy 600 treated with passivation potentials around +0.3~V to +0.4~V shows a statistical relevant decrease in material loss during active corrosion at +1.0~V.
Comparing to the amount of measured dissolved mass at $E_{(pre)cond}$ = -0.3~V, where no passivation is expected, the passivity at $E_{(pre)cond}$ = +0.3~V shows an expected reduction of the mass-loss by around 50~\%.
In the case of anodic dissolution of the metal surface after $E_{(pre)cond}$ = +0.5~V, the measured dissolved mass at +1.0~V increased again compared to passivation at +0.3~V. 
Hence, the corrosion resistance of the passive film formed at this potential is again reduced, which agrees well with the observed passivity range and the increasing passive current in the Tafel plot.

Further examination of the ratios of dissolved mass surprisingly revealed that nickel, chromium and iron were demonstrating entirely different dissolution trends depending on the passivation potential.
Specifically, \textbf{Fig. \ref{fig2}b)} shows the relative elemental ratio of dissolved mass of individual alloy elements normalized by their bulk ratio, where zero, positive and negative values indicate equivalent, more and less dissolved mass compared to the bulk composition, respectively.

Comparing the trend of dissolution for the elements, nickel exhibited a higher dissolution rate compared to its bulk ratio at all passivation potentials, with a clear reduction by about 50\% in the passive region above +0.3~V. This is also associated to the main reduction of the total mass loss observed in \textbf{Fig. \ref{fig2}a)}.

In contrast to nickel, chromium shows a weakly passivation potential-independent dissolution below bulk concentration. 
However, iron is showing a very clear passivation-dependent dissolution trend. Iron dissolved less compared to bulk ratio at lower potentials but gradually shifted into over-stoichiometric dissolution at higher passivation potential of +0.5~V.

The dissolution profile of iron shown in \textbf{Fig. \ref{fig2}c)} reveals another key feature, which may be relevant for understanding the underlying mechanism:
As indicated in the figure, the detected iron release exhibited a two-step mechanism - anodic dissolution at +1.0~V and cathodic dissolution when jumping from +1.0~V to -0.3~V, indicated by red dotted and shaded areas, respectively.

This observation suggests that the iron in the passive film undergoes at least two types of dissolution mechanisms, occurring under different polarization potentials. 
Based on Pourbaix diagrams the detected iron species at anodic polarization potential of +1.0~V may originate from the release of ferrate anion (FeO$_4^{2-}$).
The additional peak upon jumping cathodically back from +1.0~V to -0.3~V, indicates dissolution of a surface bound excess iron species formed during anodic corrosion.

In contrast to the two-step dissolution pattern of iron, nickel shows a well reproducible profile over several repetitions, which is characterised by a $\sim$1 order of magnitude dissolution rate increase, above bulk ratio, at +1.0~V.
Similar reproducibility of dissolution profiles is observed for Cr. For this element the first sequence in a repetitive set shows a delayed dissolution kinetic, which may relate to an effect of the initial native oxide, adapting to flow cell environment after the first cycle. 

\paragraph{Results of \textit{in situ} EC-AFM surface morphology probing}
\begin{figure*}
    \centering
    \includegraphics[scale=0.45]{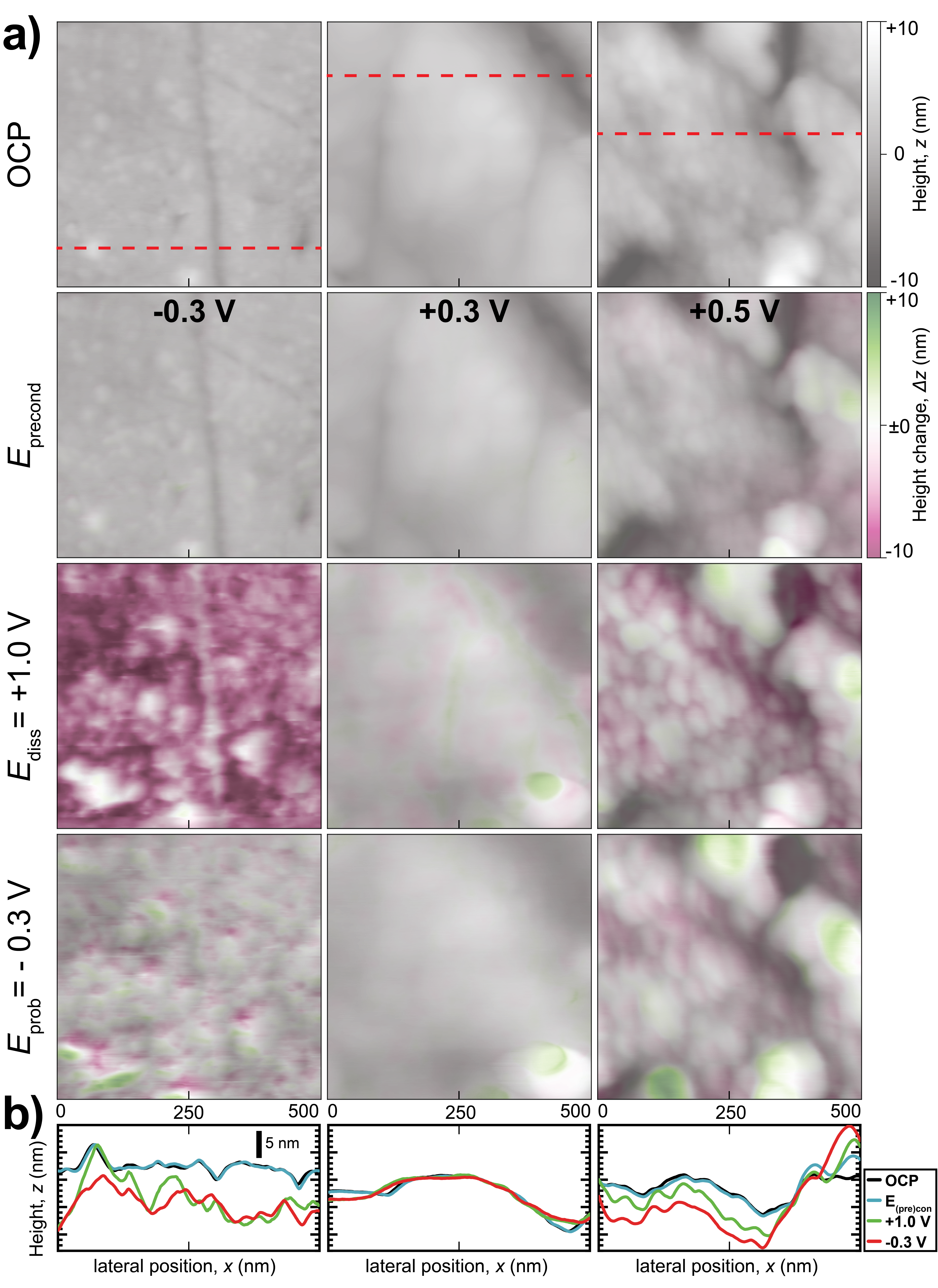}
    \caption{AFM topography during different stages ($E_{(pre)cond}$, $E_{diss}$, $E_{prob}$) of multiple step voltammetry with different (pre-)conditioning potentials $E_{(pre)cond}$ = -0.3, +0.3 and +0.5 V. 
    Red dashed lines indicate the position of the line sections shown in \textit{b)}. 
    RMS of each picture and relative change compared to the initial OCP-topography can be found in \textit{Tab. \ref{tab1}}.}
    \label{fig3}
\end{figure*}
\begin{table}[]
    \centering
    \begin{tabular}{P{2cm} P{1.6cm} P{1.6cm} P{1.6cm}}
    \hline
    Potential   & \multicolumn{3}{c}{RMS {[}nm{]} (relative to OCP)} \\ 
    $E_{precon}$ = & -0.3 V          & +0.3 V          & +0.5 V      \\ \hline
    OCP         & 0.83 (1.0)      & 2.32 (1.0)      & 3.05 (1.0)     \\ 
    precon      & 0.84 (1.0)      & 2.36 (1.0)      & 3.27 (1.1)     \\ 
    +1.0 V      & 2.37 (2.8)      & 2.57 (1.1)      & 4.12 (1.4)     \\ 
    -0.3 V      & 2.09 (2.5)      & 2.86 (1.2)      & 4.98 (1.6)     \\ \hline
    \end{tabular}
    \caption{Root mean square (RMS) values and their relative change of images shown in \textbf{Fig. \ref{fig3}}}
    \label{tab1}
\end{table}

To complement integral ICP-MS data, we further visualised the (local) evolution of the surface morphology upon passivisation using electrochemically modulated AFM measurements.

\textbf{Fig. \ref{fig3}a)} shows representative real-time changes in surface topography of alloy 600 during corrosion after different conditioning potentials $E_{(pre)cond}$ where applied (similar to ICP-MS). 
The red dashed lines indicated in the OCP (open circuit potential) topograms show line profiles that are compared in \textbf{Fig. \ref{fig3}b)}, for $E_{(pre)cond}$ (blue), +1.0~V (green) and -0.3~V (red), respectively.

As a side note, for better visualisation height changes were overlaid on the recorded images. Increase in height is indicated by green shading, while dissolution from the surface is shown in magenta.
Further, AFM topography does not provide information of absolute but only relative heights. 
Hence, consecutively measured topographies were shifted during post-processing, using a referencing protocol described in the methods section, in order to approximate the absolute change in topography during corrosion.
The selected images summarise recurring patterns and characteristics seen in repeat experiments.

The potential-dependent EC-AFM images reveal a number of interesting details as described in the following. 

First, during the passive film formation (preconditioning) the surface morphology of all three examined precondition potentials show no significant changes compared to OCP. 
Only very minor height variations are observed. 
Jumping from precondition potential to dissolution potential, the alloy treated at $E_{(pre)cond}$ = -0.3~V shows a very vigorous dissolution. 
The RMS roughness increases significantly (see Table \ref{tab1}), and most of the pronounced features (polishing scratches, etc.) become indistinguishable, while only few areas remained stable for referencing the corrosive height loss.

Second, and in a sharp contrast, the surface morphology during anodic corrosion has no significant change when the passive film was grown under polarization of +0.3~V. 
Obviously, the surface oxide film does protect the surface very effectively, and dissolution occurs homogeneously along preexisting features such as scratches.

Third, surfaces passivated at +0.5~V again show more pronounced corrosive degradation at +1.0~V. 
Here, the surface morphology did not undergo a similarly massive dissolution seen for preconditioning at -0.3~V. 
However, significant dissolution preferentially occurred along the edges of the nano-granular features. Additionally, in some areas measurable material growth was observed, indicating possible local re-precipitation reactions.

Fourth, upon switching the polarization from +1.0~V to -0.3~V, the dissolution reactions stopped and only minor topographic changes were observed for preconditioning at +0.3~V and +0.5~V. 
However, for samples pretreated at -0.3~V dissolution was still observed at the top of particles.
This correlates well with the observed significant cathodic iron dissolution peak (see \textbf{Fig. \ref{fig2}c)}), which further supports that the surface enriches in iron oxide during active corrosion, if corrosion starts without a significant passive film presence. 

\paragraph{Overall discussion of results}
In summary, the AFM topography analysis agrees very well with the ICP-MS online analysis results, which reveals that the passivation potential, i.e. the conditions during passive film formation are directly influencing the behaviour during active corrosion. 
Together with the transition of the ICP-MS dissolution pattern our data suggests that the quantitative elemental dissolution during active corrosion is highly dependent on the passive film properties.
In the following, we want to discuss several interesting aspects of our data.

First, if the material is directly corroded after polarization at -0.3~V, ICP-MS and AFM data suggested a massive material loss. 
Given the Pourbaix diagram, one may expect a pure chromium oxide layer, as the most stable chromium phase is chromium oxide at -0.3~V at the experimental pH. 
Further, the passive film might be much thinner, or even defect rich, or not complete. 
As a consequence, any protective effect is limited or not even present.
Second, in line with this thought we find clear and pronounced cathodic iron dissolution only after corrosion starting directly from -0.3~V. 
This suggests that a steady state surface film evolves during anodic breakdown, which is significantly enriched in iron, and presumably iron oxide. 
This is consistent with the Pourbaix diagram suggesting that  Chromium oxide is not stable at corrosive conditions at 1.0~V. 
Our data may hence indicate that iron oxide forms as a transient oxide during corrosion, which is specifically growing at defects or areas that are not covered by a stable chromium oxide.
Further, during corrosion, this oxide is not sufficient to suppress the material loss as effectively as a proper passive film grown at passivation potentials within the passive region.
This iron oxide enrichment is evident form the significant iron dissolution peak during repassivation, after corrosion at 1.0~V. 

Third, and along the same line, passivation at $\geq$ +0.3~V establishes a passive film, which can significantly decrease the material loss during active corrosion. 
Still, we find a small but detectable cathodic iron peak after active corrosion. 
This may suggest that the passive film breaks down locally, presumably at defect rich sites, and/or iron oxide accumulates at the outside of the passivating chromium oxide.

Considering the AFM data at +0.3~V, which shows a rather uniform material loss at defect rich areas (i.e. preexisting polishing defects, etc.), our findings appear more consistent with an increased iron dissolution across the defect rich areas of the passive film. 
Further, as seen in \textbf{Fig. \ref{fig2}b)} relative iron dissolution also increases significantly if a passive film is grown between +0.3-0.5~V. 
Hence, iron may accumulate at the oxide|water interface and/or gets transported through the passive film at a higher ratio during corrosive dissolution. This suggests a change in the passive film chemistry. 

Fourth, after passivation at +0.5~V the again increasing iron content during cathodic polarization after corrosion, suggests a decisive role of defect levels on the transport across the passive film during corrosion. 
As such, the growth conditions apparently render the passive film with different defect levels. 
Together with the increase in iron ratio during corrosion (see again \textbf{Fig. \ref{fig2}b)}) this suggest a structural change in the oxide that accelerated iron diffusion across the existing passive film in a high-field condition applied during corrosion at +1.0~V.
This is further supported by AFM, where we see increased dissolution at areas that are defect rich. 
The high field model (HFM)\cite{Cabrera.1949,tranchida2018electronic}  describing the built up of the passive film during $E_{(pre)cond}$ provides an explanation for the differences in density of defects, and hence, corrosion resistance. 

Also, during the cathodic jump AFM data at +0.5~V shows further material loss at the preexisting grain boundaries. 
The passive film grown at +0.5~V may hence trigger faster oxide growth, due to an increase of defect rich areas, as indicated by the localised loss seen in AFM topography. 

We refrain from interpreting too much into the apparent material growth areas at this point in time, as AFM is not an absolute technique, and post-processing shifts may not truly reflect any growth areas, while material loss is clearly distinguishable. 
Along this line, one important aspect to consider when comparing the processes observed in both the AFM and the ICP-MS flow cell is their general differences:
While the ICP-MS requires a constant flow of electrolyte the AFM measurements are conducted in a stagnant solution. 
Quantitative comparision also indicates that corrosion is more severe in AFM, where for instance  after preconditioning at +0.3~V a few nm of material are lost during corrosion at 1.0~V, while ICP-MS indicates only a monolayer material loss per cycle (300 ng/cm² $\cong$\ 0.35 nm). 
This discrepancy is considerably lower when the material corrodes less. 

Hence, the dissolution in AFM is also accelerated due to typical surface confinement in an AFM (typically the AFM cantilever holder is separated by less than 200 $\mu$m from the probed surfaces), which renders this experiment more effectively "crevice like", where electrolyte of the dissolving species may render the interface at a different pH. This needs to be considered for AFM and scanning probe techniques in general when studying corrosion, and may become a feature, if properly controlled. 
How scanning probe conditions can even compare to flow-cell experiments, and other electrochemical cells, is in our view an aspect that will need further attention, e.g. by directly coupling AFM and ICP-MS, and by further comparative studies with model systems. 

\section{Conclusions}
In this work, we successfully used two complementary analytical techniques to study the influence of passivation potential on the formed metal oxide film quality in real-time.
We designed an AFM cell that performs very well during corrosion. The - for scanning probe techniques - novel plate-plate geometry provides an idealized field distribution. 
Using an ICP-MS flow cell and EC-AFM, passivation and anodic dissolution was tracked with both time resolved elemental dissolution rates during chronoamperometric pulse experiments and complementary changes in morphology, respectively.
Regarding the characterised passivation potentials on alloy 600 we could find an optimum at around +0.3 - +0.4~V in neutral NaCl solution exhibiting lowest dissolution and therefore forming the most stable passive film during anodic corrosion at +1.0~V.

Our results suggest the following specific conclusions:
\begin{itemize}
    \item Quality of the passive film depends on the potential at which it is grown.
    \item The passivation potential controls the defect density and ion-conducting properties of the passive film during active corrosion.
    \item Combination of \textit{in situ} techniques provide information about stability, indirectly composition and local stability of the passive film.
\end{itemize}

With the ICP-MS flow cell and the EC-AFM it is hence possible to relate two \textit{in situ} techniques to consistently describe the quality and stability of a passive film. 
In combination both techniques provide real-time and localized information under operating conditions, and are much closer to realistic corrosion conditions, compared to other surface sensitive analytical methods.

\section*{Acknowledgements}
The authors acknowledge support by the European Research Council (Grant: CSI.interface, ERC-StG 677663, characterization of surfaces). 

\section*{Competing Interests}
The Authors declare no Competing Financial or Non-Financial Interests. 

\section*{Data availability}
The raw and processed data required to reproduce these findings are available from the corresponding author via www.repositum.tuwien.ac.at upon reasonable request.

\section*{Methods and Materials}
\paragraph{Chemicals and Materials.} Sodium chloride (Carl Roth, p.a.) and Milli-Q water (resistivity \textgreater 18 M$\Omega\cdot$ cm, total organic carbon < 4 ppb) was used for making electrolyte solutions. Alloy 600 was obtained from VDM-Metals, Cr5 was provided by Hauke Springer (MPI f. Eisenforschung, D\"usseldorf). 
The metals were ground with sand paper of decreasing grain size (from P80 to P2500), then polished with diamond paste down to 0.05 $\mu$m.
Prior to electrochemical experiments the metals were maximum 5 min \textit{in situ} in the ICP-MS flow cell by 5 minutes potentiostatic polarization at $\rm-0.2 V$ vs OCP.

\paragraph{Atomic Force Microscopy (AFM)}
AFM topographies were taken on with AM-mode Cypher ES (Asylum Research, Oxford Instruments, Santa Barbara, CA) using Arrow\textsuperscript{TM} UHFAuD (NanoWorld, CH) and SCOUT 350 RAu (NuNano, UK) probes. blueDrive was used to oscillate the probes. Embedded NBA samples were contacted with silver glue. 
For embedded samples a rectangular piece of metal of about 5 mm side length was used. The block material was embedded in methylmethacrylate based VariDur 200 (Buehler). A PalmSens 4 potentiostat (PalmSens, Netherlands) was used.
\subparagraph*{Image processing}
Prior drift correction and further analysis images were only processed by applying flattening of first order. For alignment in z-direction a minimum of 10 line profiles were chosen per time-series and adapted in height to match significant - and non-changing - features of the images, e.g. deep scratches from polishing or distinct peak features that remain over time.

\paragraph{Cell design of the electrochemical AFM (EC-AFM)}
Our home-built electrochemical cell design (see schematics in \textbf{Fig. \ref{fig1} b) \& c)} is inspired by concepts of Valtiner et al.\cite{Valtiner.2011} and derives from a standard Perfusion Cantilever Holder (Asylum Research). We modified it by placing a platinum foil as a counter electrode (CE) at the bottom side of the cantilever holder. When assembled the working electrode (WE) and the platinum CE form a parallel plate geometry resulting in a very evenly distributed electrical field. A majority of the upper surface is covered with the platinum foil to provide a high area, hence avoiding any rate limitation from reactions at the CE. Only the optical path for the laser of the AFM and a feed trough to connect the reference electrode remain cleared. A commercial silver/silver chloride reference electrode is connected with the cell via a capillary. 
For compensation of the expected high electrical current especially during anodic dissolution and given that the CE is in close vicinity to the sample a reduction of the WE size is limiting the possibilty of disturbance through reactions happening at the CE (e.g. hydrogen evolution and bubble formation).
The size ration of WE to CE is at least 5:1. 

\paragraph{Inductively Coupled Plasma Mass Spectrometry (ICP-MS)}
Measurements were performed using an Agilent 7900 ICP-MS from Agilent Technologies. Calibration was performed with multi-element standard solutions provided by Agilent and Inorganic Ventures. The ICP-MS uses a collision cell with 5~mL/min flow of helium as cell gas. Downstream of the electrochemical cell the analyte was mixed with standard solution. Cobalt was chosen as internal standard due to its similar mass with most of the alloy components. Electrochemical experiments were conducted in a home-built flow cell out of PEEK and PTFE insipired by the Ogle- and Mayrhofer designs. \cite{ogle_anodic_2000,schuppert_scanning_2012,klemm_coupling_2011} 
The exposed electrode area is circular and sealed with a 3 mm inner diameter O-ring. 
Inset in \textbf{Figure \ref{fig1}a)} shows schematically the flow-cell used in this work. Pumping with pressurized nitrogen establishes a stable and pulsation free laminar flow of electrolyte. Flow was monitored both by an in-flow pressure sensor and by weighing the collected waste electrolyte after ICP-MS.
The flow is set to $6 \pm 0.2$ mg of solution per second. 
Before each experiment the electrolyte was purged with compressed and filtered air for at least 30 minutes to guarantee the same concentration of dissolved oxygen. 
A Biologic VSP-300 potentiostat (Biologic, France) was used.
All electrochemical experiments were done using a Ag|AgCl-Electrode as reference and presented data is referenced to that potential.

\section*{References}
\bibliography{Library}

\begin{thebibliography}{10}

\bibitem{Lu.2007b}
B.~T. Lu, J.~L. Luo, and Y.~C. Lu, ``{A Mechanistic Study on Lead-Induced
  Passivity-Degradation of Nickel-Based Alloy},'' {\em {Journal of The
  Electrochemical Society}}, vol.~154, no.~8, p.~C379, 2007.

\bibitem{rosenberg_nickel_1968}
S.~J. Rosenberg, {\em Nickel and its alloys}.
\newblock National Bureau of Standards, 1968.

\bibitem{rowcliffe_perspectives_2009}
A.~Rowcliffe, L.~Mansur, D.~Hoelzer, and R.~Nanstad, ``Perspectives on
  radiation effects in nickel-base alloys for applications in advanced
  reactors,'' {\em Journal of Nuclear Materials}, vol.~392, pp.~341--352, July
  2009.

\bibitem{henderson_nickel_2004}
M.~B. Henderson, D.~Arrell, R.~Larsson, M.~Heobel, and G.~Marchant, ``Nickel
  based superalloy welding practices for industrial gas turbine applications,''
  {\em Science and Technology of Welding and Joining}, vol.~9, pp.~13--21, Feb.
  2004.

\bibitem{persaud_characterization_2018}
S.~Persaud, B.~Langelier, A.~Korinek, S.~Ramamurthy, G.~Botton, and R.~Newman,
  ``Characterization of initial intergranular oxidation processes in alloy 600
  at a sub-nanometer scale,'' {\em Corrosion Science}, vol.~133, pp.~36--47,
  Apr. 2018.

\bibitem{Hayes.2006}
J.~R. Hayes, J.~J. Gray, A.~W. Szmodis, and C.~A. Orme, ``{Influence of
  Chromium and Molybdenum on the Corrosion of Nickel-Based Alloys},'' {\em
  {CORROSION}}, vol.~62, no.~6, pp.~491--500, 2006.

\bibitem{Lloyd.2004}
A.~C. Lloyd, J.~J. No{\"e}l, S.~McIntyre, and D.~W. Shoesmith, ``{Cr, Mo and W
  alloying additions in Ni and their effect on passivity},'' {\em
  {Electrochimica Acta}}, vol.~49, no.~17-18, pp.~3015--3027, 2004.

\bibitem{ebrahimi_role_2015}
N.~Ebrahimi, P.~Jakupi, J.~No\"{e}l, and D.~Shoesmith, ``The {Role} of
  {Alloying} {Elements} on the {Crevice} {Corrosion} {Behavior} of
  {Ni}-{Cr}-{Mo} {Alloys},'' {\em CORROSION}, vol.~71, pp.~1441--1451, Sept.
  2015.

\bibitem{persaud_review_2016}
S.~Persaud and R.~Newman, ``A {Review} of {Oxidation} {Phenomena} in {Ni}
  {Alloys} {Exposed} to {Hydrogenated} {Steam} {Below} 500$^{\circ}${C},'' {\em
  Corrosion}, vol.~72, pp.~881--896, July 2016.

\bibitem{Maurice.2018}
V.~Maurice and P.~Marcus, ``{Current developments of nanoscale insight into
  corrosion protection by passive oxide films},'' {\em {Current Opinion in
  Solid State and Materials Science}}, vol.~22, no.~4, pp.~156--167, 2018.

\bibitem{Galtayries.2006}
A.~Galtayries, A.~Machet, P.~Jolivet, P.~Scott, M.~Foucault, P.~Combrade, and
  P.~Marcus, ``{Kinetics of passivation of nickel-base alloys (Alloy 600 and
  Alloy 690) in high temperature water},'' in {\em {Passivation of metals and
  semiconductors, and properties of thin oxide layers}} (P.~Marcus and
  V.~Maurice, eds.), pp.~403--409, Amsterdam and Oxford: Elsevier, 2006.

\bibitem{Marcelin.2016}
S.~Marcelin, B.~Ter-Ovanessian, and B.~Normand, ``{Electronic properties of
  passive films from the multi-frequency Mott--Schottky and power-law coupled
  approach},'' {\em {Electrochemistry Communications}}, vol.~66, pp.~62--65,
  2016.

\bibitem{Mito.2009}
Y.~Mito, M.~Ueda, and T.~Ohtsuka, ``{Photo-luminescence from passive oxide
  films on nickel and chromium by photo-excitation of UV light},'' {\em
  {Corrosion Science}}, vol.~51, no.~7, pp.~1540--1544, 2009.

\bibitem{Tsuchiya.2002}
H.~Tsuchiya, S.~Fujimoto, O.~Chihara, and T.~Shibata, ``{Semiconductive
  behavior of passive films formed on pure Cr and Fe--Cr alloys in sulfuric
  acid solution},'' {\em {Electrochimica Acta}}, vol.~47, no.~27,
  pp.~4357--4366, 2002.

\bibitem{Merola.2019}
C.~Merola, H.-W. Cheng, D.~Dworschak, C.-S. Ku, C.-Y. Chiang, F.~U. Renner, and
  M.~Valtiner, ``{Nanometer Resolved Real Time Visualization of Acidification
  and Material Breakdown in Confinement},'' {\em {Advanced Materials
  Interfaces}}, vol.~6, no.~10, p.~1802069, 2019.

\bibitem{Frankel.2017}
G.~S. Frankel, T.~Li, and J.~R. Scully, ``{Perspective---Localized Corrosion:
  Passive Film Breakdown vs Pit Growth Stability},'' {\em {Journal of The
  Electrochemical Society}}, vol.~164, no.~4, pp.~C180--C181, 2017.

\bibitem{Cabrera.1949}
N.~Cabrera and N.~F. Mott, ``{Theory of the oxidation of metals},'' {\em
  {Reports on Progress in Physics}}, vol.~12, no.~1, pp.~163--184, 1949.

\bibitem{Chao.1981}
C.~Y. Chao, L.~F. Lin, and D.~D. Macdonald, ``{A Point Defect Model for Anodic
  Passive Films: I . Film Growth Kinetics},'' {\em {Journal of The
  Electrochemical Society}}, vol.~128, no.~6, pp.~1187--1194, 1981.

\bibitem{Marcus.2012}
P.~Marcus, {\em {Corrosion mechanisms in theory and practice}}, vol.~26 of {\em
  {Corrosion technology}}.
\newblock Boca Raton: {CRC Press}, 3rd ed.~ed., 2012.

\bibitem{tranchida2018electronic}
G.~Tranchida, M.~Clesi, F.~Di~Franco, F.~Di~Quarto, and M.~Santamaria,
  ``Electronic properties and corrosion resistance of passive films on
  austenitic and duplex stainless steels,'' {\em Electrochimica Acta},
  vol.~273, pp.~412--423, 2018.

\bibitem{Chen.2020}
H.~Chen, Z.~Qin, M.~He, Y.~Liu, and Z.~Wu, ``{Application of Electrochemical
  Atomic Force Microscopy (EC-AFM) in the Corrosion Study of Metallic
  Materials},'' {\em {Materials}}, vol.~13, no.~3, 2020.

\bibitem{Shi2018}
Y.~Shi, L.~Collins, N.~Balke, P.~Liaw, and B.~Yang, ``In-situ
  electrochemical-afm study of localized corrosion of al$_x$cocrfeni
  high-entropy alloys in chloride solution,'' {\em Applied Surface Science},
  vol.~439, 05 2018.

\bibitem{Deng2018}
X.~Deng, F.~Galli, and M.~Koper, ``In situ electrochemical afm imaging of a pt
  electrode in sulfuric acid under potential cycling conditions,'' {\em Journal
  of the American Chemical Society}, vol.~140, 09 2018.

\bibitem{Maurice.2007}
V.~Maurice, T.~Nakamura, L.~Klein, and P.~Marcus, ``{7 - Initial stages of
  localised corrosion by pitting of passivated nickel surfaces studied by
  scanning tunnelling microscopy (STM) and atomic force microscopy (AFM)},'' in
  {\em {Local Probe Techniques for Corrosion Research}} (R.~Oltra, V.~Maurice,
  R.~Akid, and P.~Marcus, eds.), {European Federation of Corrosion (EFC)
  Series}, pp.~71--83, {Woodhead Publishing}, 2007.

\bibitem{beverskog_revised_1996}
B.~Beverskog and I.~Puigdomenech, ``Revised pourbaix diagrams for iron at
  {25}-{300} $^{\circ}${C},'' {\em Corrosion Science}, vol.~38, pp.~2121--2135,
  Dec. 1996.

\bibitem{Valtiner.2011}
M.~Valtiner, G.~N. Ankah, A.~Bashir, and F.~U. Renner, ``{Atomic force
  microscope imaging and force measurements at electrified and actively
  corroding interfaces: challenges and novel cell design},'' {\em {The Review
  of scientific instruments}}, vol.~82, no.~2, p.~023703, 2011.

\bibitem{ogle_anodic_2000}
K.~Ogle and S.~Weber, ``Anodic dissolution of 304 stainless steel using atomic
  emission spectroelectrochemistry,'' {\em Journal of The Electrochemical
  Society}, vol.~147, no.~5, pp.~1770--1780, 2000.

\bibitem{schuppert_scanning_2012}
A.~K. Schuppert, A.~A. Topalov, I.~Katsounaros, S.~O. Klemm, and K.~J.~J.
  Mayrhofer, ``A {Scanning} {Flow} {Cell} {System} for {Fully} {Automated}
  {Screening} of {Electrocatalyst} {Materials},'' {\em Journal of The
  Electrochemical Society}, vol.~159, no.~11, pp.~F670--F675, 2012.

\bibitem{klemm_coupling_2011}
S.~O. Klemm, A.~A. Topalov, C.~A. Laska, and K.~J. Mayrhofer, ``Coupling of a
  high throughput microelectrochemical cell with online multielemental trace
  analysis by {ICP}-{MS},'' {\em Electrochemistry Communications}, vol.~13,
  pp.~1533--1535, Dec. 2011.

\end{thebibliography}
\bibliographystyle{ieeetr}

\end{document}